\documentstyle[preprint,aps,tighten]{revtex}
\newcommand{\beq}{\begin{equation}}
\newcommand{\eeq}{\end{equation}}
\newcommand{\beqa}{\begin{eqnarray}}
\newcommand{\eeqa}{\end{eqnarray}}
\def\ket#1{| #1 \rangle}
\def\vev#1{\langle #1 \rangle}
\def\a{\alpha}
\def\b{\beta}
\def\l{\ell}
\def\la{\lambda}

\def\c{\cos\theta_{ms}}
\def\s{\sin\theta_{ms}}
\def\cp{\cos\theta_{md}}
\def\sp{\sin\theta_{md}}
\def\H{{\cal H}}
\def\tt{\tilde{\tau}_R}

\begin{document}
\draft
{
\preprint{\vbox{\hbox{WIS-95/30/Jul-PH}
                \hbox{hep-ph/9507344}
                \hbox{July, 1995} }}
\title{Non-Standard Neutrino Interactions and Neutrino
Oscillation Experiments}
\author{Yuval Grossman}
\address{ \vbox{\vskip 0.5truecm}
  Department of Particle Physics \\
      Weizmann Institute of Science, Rehovot 76100, Israel}
\maketitle
\begin{abstract}%
In analyzing neutrino oscillation experiments
it is often assumed that while
new physics contributes to neutrino masses,
neutrino interactions are given by the Standard Model.
We develop a formalism to study new physics effects
in neutrino interactions using oscillation experiments.
We argue that the notion of
branching ratio is not appropriate in this case.
We show that a neutrino appearance experiment with sensitivity
to oscillation probability $P_{ij}^{exp}$ can
detect new physics in neutrino interactions
if its strength $G_N$ satisfies $(G_N/G_F)^2 \sim P_{ij}^{exp}$.
Using our formalism we show
how current experiments on neutrino oscillation give
bounds on the new interactions in various new physics scenarios.
\end{abstract}
}

\newpage
\section{Introduction}
The goal of neutrino oscillation experiments
is to probe those extensions of the Standard Model (SM)
which predict non-vanishing neutrino masses.
However, in the usual treatment of neutrino oscillation experiments, it is
often assumed that neutrino interactions are
described just by the SM \cite{rev}.
While we know that this is a good approximation,
often physics beyond the SM induces also new neutrino interactions.
If New Physics (NP) contributes significantly to neutrino interactions,
the conclusions that we draw from the experimental data can be affected.
For example, even for massless neutrinos,
the NP  can allow for weak eigenstate
muon neutrino to produce an electron in the detector, and in this case,
we may erroneously conclude that oscillations have occurred.

To search for NP effects in massive fermion interactions
(quarks and charged leptons) the experimentally measured branching ratios
are used in a straightforward way.
However, in the case of neutrinos,
there are two important subtleties:
\begin{itemize}
\item{Neutrino masses are unknown, and their difference may be very
small. In such a case,
experiments cannot observe neutrinos as mass eigenstates,
and the results are sensitive to the time evolution of the
flavor eigenstates.}
\item{The neutrino flavor is identified by
charged current interactions. Since
NP may modify them, the identification of the neutrinos cannot be done
in a model independent way.}
\end{itemize}
The results of neutrino oscillation experiments are
sensitive to the following three ingredients: The production process,
the time evolution and the detection process.
It is impossible to separate the
NP contributions to the neutrino production or detection process:
a formalism that combines all the three ingredients is necessary.

\section{Formalism}
For simplicity, and without loss of generality,
we assume two neutrino flavors, CP conservation and that
the neutrinos are highly relativistic.
First, we define the bases we use.
Since the mass basis is well defined,
we express all neutrino states
as superpositions of mass eigenstates.
We always work in the mass basis for the charged leptons.
Then, the weak interactions define the weak basis where the
neutrino weak eigenstates are $SU(2)$ partners of the
charged leptons.
We start with a specific example and
generalize it later. We consider a
muon neutrino beam
produced by $\pi \to \mu \nu$ decay,
and the subsequent detection of electron neutrinos through
$\nu n\to ep$.
The weak eigenstate $\ket{\nu_\mu^W}$, is given
by a superposition of mass eigenstates $\ket{\nu_\a^m}$ as
\beq \label{defw}
\ket{\nu_\mu^W} = \sum_\a U^W_{\mu\a}\ket{\nu_\a^m},
\eeq
so that $|U^W_{\mu\a}|^2 \propto |\vev{\nu_\a^m,\mu|\H^W|\pi}|^2$ where
$\H^W$ is the weak interaction Hamiltonian.
In the presence of NP there might be
extra carriers of the charged current interaction besides the $W$ boson.
Therefore, the neutrino produced by $\pi \to \mu \nu$
may be different from $\ket{\nu_\mu^W}$.
We define this neutrino as a source
basis eigenstate $\ket{\nu_\mu^s}$,  given by
a different superposition of mass eigenstates
\beq \label{defs}
\ket{\nu_\mu^s} = \sum_\a U^s_{\mu\a} \ket{\nu_\a^m},
\eeq
so that $|U^s_{\mu\a}|^2 \propto |\vev{\nu_\a^m,\mu|\H|\pi}|^2$,
$\H=\H^W+\H^{NP}$ where $\H^{NP}$ is the NP interaction Hamiltonian.
Similarly, we define the neutrino detected by $\nu n\to ep$
as a detector
basis eigenstate $\ket{\nu_e^d}$,  given
by another superposition of mass eigenstates
\beq \label{defd}
\ket{\nu_e^d} = \sum_\a U^d_{e\a}\ket{\nu_\a^m},
\eeq
so that $|U^d_{e\a}|^2 \propto |\vev{\nu_\a^m,n|\H|e,p}|^2$.

In general, for any neutrino oscillation experiment
it is useful to use the following bases:
\begin{itemize}
\item{%
The mass basis, $\left\{\ket{\nu_\a^m}\right\}$, where the
neutrino mass matrix is diagonal.
}
\item{%
The weak basis, $\left\{\ket{\nu_\a^W}\right\}$,
where the leptonic couplings of the $W$ are diagonal.
}
\item{%
The source basis, $\left\{\ket{\nu_\a^s}\right\}$,
where the interaction of the production process is diagonal.
}
\item{%
The detector basis, $\left\{\ket{\nu_\a^d}\right\}$,
where the interaction of the detection process is diagonal.
}
\end{itemize}
When neutrino interactions are fully described by the SM,
the last three definitions
coincide, and this basis is usually called the interaction
or the flavor basis. However, the main
lesson from the above discussion is that in the presence of NP
those three bases can be different.

The source and the detector bases are related to the mass basis
through the unitarity transformations
\beq \label{defbasis}
\ket{\nu_\l^s}=\sum_{\a} U^s_{\l\a} \ket{\nu_\a^m},\ \ \ \
\ket{\nu_\l^d}=\sum_{\a} U^d_{\l\a} \ket{\nu_\a^m},
\eeq
with $\l=e,\mu,\tau$.
The amplitude for finding a $\nu_n^d$ in the original $\nu_\l^s$ beam
at time $t$ is
\beq \label{amp}
\vev{\nu_n^d|\nu_\l^s}(t)=
\sum_{\a,\b} \vev{\nu_\b^m|U^{d\dagger}_{\b n} e^{-iE_\a t}
U^s_{\l\a}|\nu_a^m}=
\sum_{\a} e^{-iE_\a t} U^s_{\l\a} U^{d*}_{n\a}\,,
\eeq
where in the last step we have used the orthogonality of the mass eigenstates.
The probability of finding a $\nu_n^d$ in the original $\nu_\l^s$
beam at time $t$ is
\beqa \label{rate}
P_{n\l}(t)&=&
\left|\vev{\nu_n^d|\nu_\l^s}(t)\right|^2=
\left(\sum_{\a} e^{-iE_\a t} U^s_{\l\a} U^{d*}_{n\a}\right)
\left(\sum_{\b} e^{iE_\b t} U^{s*}_{\l\b} U^d_{n\b}\right) \\
&=& \sum_{\a,\b} \left|U^s_{\l\a} U^{s*}_{\l\b} U^{d*}_{n\a} U^d_{n\b} \right|
\cos\left[(E_\a-E_\b)t-
\arg\left(U^s_{\l\a} U^{s*}_{\l\b} U^{d*}_{\a n}
U^d_{\b n}\right)\right]. \nonumber
\eeqa
For two neutrino flavors and with CP conservation we have
\beq \label{defmat}
U=\pmatrix{\c&-\s\cr \s&\c\cr},\ \ \ \
V=\pmatrix{\cp&-\sp\cr \sp&\cp\cr},
\eeq
with $|\theta_{ms}|,|\theta_{md}|\leq \pi/ 4$.
Define
\beq \label{defx}
x\equiv {\Delta m^2 t \over 4 E},\ \ \ \
\Delta m^2 \equiv m_1^2-m_2^2,\ \ \ \ \
\theta_{sd}\equiv \theta_{md}-\theta_{ms}\,.
\eeq
Using
$E_\a-E_\b \approx (m_\a^2-m_\b^2)/2E$,
we get our main result:
\beq \label{simsec}
P_{e\mu}(x)=\sin^2\theta_{sd} + \sin2\theta_{md} \,
\sin2\theta_{ms} \,\sin^2x.
\eeq
Few points are in order:

\begin{enumerate}
\item{When neutrino interactions are described by the SM,
$\theta_{md}=\theta_{ms}=\theta$,
and Eq.(\ref{simsec}) reduces to the known result,
$P_{e\mu}(x)=\sin^2 2\theta \sin^2x$ \cite{rev}.}

\item{NP that affects the production and the detection
processes in the same way
cannot be detected in appearance experiments.
In those cases the flavor eigenstates are the same for all processes,
even if they may differ from the
weak eigenstates. Then we have, $\theta_{md}=\theta_{ms}$ and Eq.(\ref{simsec})
reduces again to the standard form, so that
we cannot distinguish
this situation from the SM interaction case.
}

\item{%
Experiments are working with neutrino beams, not necessarily monoenergetic.
$P_{e\mu}^{exp}$ is defined to be the total
appearance probability of a specific
experiment, where the dependence on the energy spectrum of the beam
and on the baseline
length $L$, is included.
If neutrinos are produced by several
decays of different initial states, then
\beq \label{sevd}
P_{e\mu}^{exp}=\sum_i a_i \tilde{P}_{e\mu}^{exp}(i),
\eeq
where $a_i$ is the relative weight of the $i$'th decay
mode in the neutrino beam,
and $\tilde{P}_{e\mu}^{exp}(i)$ is the appearance probability had only the
$i$'th decay mode been responsible for the neutrino production.}

\item{In two limits, $\Delta m^2 \gg E/L$ ($x \to \infty$) and
$\Delta m^2=0$ ($x = 0$), $P_{e\mu}$ is $x$ independent.
When $x$ is large, $\sin^2x$ averages to $1 \over 2$,
then Eq.(\ref{simsec}) gives
\beq \label{pbr}
P_{e\mu} =
\sin^2\theta_{sd}+ {1\over2}\sin2\theta_{ms} \, \sin2\theta_{md}\,.
\eeq
For massless (or degenerate) neutrinos, $x=0$
and the appearance probability becomes
\beq \label{npimp}
P_{e\mu} = \sin^2\theta_{sd}.
\eeq
We learn that
a signal can be seen in appearance experiments
even for massless neutrinos. This is the case when
$\theta_{sd} \ne 0$, namely, when the interaction in the
production process
is different from the interaction of the
detection process.
This signal is constant in distance, and does not have
an oscillation pattern.
We conclude: {\it
a distance-independent signal
is not enough to prove that neutrinos are massive.
Only an oscillation pattern provides a proof.}
}
\end{enumerate}

Experimentally, we know that neutrino interactions are dominantly
those of the SM.
Therefore, while  $\theta_{ms}$ and $\theta_{md}$
may be large, their difference has to be small.
It is therefore reasonable to
work in the weak basis
and treat the NP as a perturbation.
In the two generation case we define two small angles,
$\theta_{Ws}$ and $\theta_{Wd}$, that parameterize the
deviation from the weak basis, and a third angle
(not necessarily small) $\theta_{Wm}$,
that rotates from the weak to the mass basis.
Using
$\theta_{ab}+\theta_{bc}=\theta_{ac}$ and $\theta_{ab}=-\theta_{ba}$,
we get the relations between these angles
and those defined in (\ref{defmat}):
$\theta_{sd} = \theta_{Wd}-\theta_{Ws}$ and
$\theta_{md} = \theta_{Wd}-\theta_{Wm}$.

To find the rotation angles we use the previously mentioned example,
but the final result is general. We consider a
muon neutrino beam
produced by $\pi \to \mu \nu$ decay,
while the subsequent electron neutrinos are detected through
$\nu n\to ep$.
For SM interactions the produced neutrinos are $\nu_\mu^W$, and
the detected ones are $\nu_e^W$.
We are interested in NP that gives
effective couplings of the form $G_N^s \bar{u} d \bar{\mu} \nu_e^W$ and
$G_N^d \bar{u} d \bar{e} \nu_\mu^W$ with $|G_N^s| \neq |G_N^d|$.
Then, the produced and detected neutrinos are superpositions
of weak eigenstates, $\nu_\mu^s \sim G_F\nu_\mu^W + G_N^s\nu_e^W$
and $\nu_e^d \sim G_F\nu_e^W + G_N^d\nu_\mu^W$,
and it follows that
\beq \label{rotws}
\theta_{Ws} \sim {G_N^s \over G_F},\ \ \ \
\theta_{Wd} \sim {G_N^d \over G_F}.
\eeq
It is useful to express the appearance probability in terms of the
rotation angle from the weak to the mass basis, $\theta_{Wm}$,
and the NP strength, $G_N^s$ and $G_N^d$.
For $x=0$ we get from (\ref{npimp})
\beq \label{xzero}
P_{e\mu} \approx \left({G_N^s-G_N^d \over G_F}\right)^2.
\eeq
Experimentally we know that in the $x \to \infty$ limit
all the relevant angles are small.
Then we get from (\ref{pbr})
\beq \label{xinfty}
P_{e\mu} \approx \left({G_N^s \over G_F}\right)^2 +
\left({G_N^d \over G_F}\right)^2 +
2 \theta_{Wm}\left({G_N^s+G_N^d \over G_F}\right) + 2 \theta_{Wm}^2.
\eeq

{}From the above formulae we can
obtain an order of magnitude estimate of the strength of
the experimentally relevant NP.
An experiment with
sensitivity to oscillation probability
$P_{e\mu}^{exp}$ can probe NP in neutrino interactions
when its effective strength $G_N \sim {\rm max}(G_N^s,G_N^d)$
satisfies
\beq \label{npimpf}
\left({G_N\over G_F}\right)^2 \gtrsim P_{e\mu}^{exp}\,.
\eeq

\section{Branching Ratio}
Measurements of Branching Ratios (BR) are widely used in
searching for NP effects.
The meaning of BR is unambiguous when discussing
quarks and charged leptons, for which a BR measures the
transition rate between mass eigenstates.
The main advantage of using BR  is that
only the production process is relevant to the calculation,
and one need not worry how the decay products
are detected.
Therefore, measuring a BR has to be
independent of the experimental setup and of the theoretical model
under study.
Since experiments cannot detect neutrinos
as mass eigenstates, for neutrinos
these requirements are not satisfied:
the time evolution of the flavor eigenstates,
and the NP effects in the detection process cannot, in general,
be separated from the analysis of the experimental results.
Therefore, the extension of the BR notion to the neutrino case is
problematic.

We see three major disadvantages in using  BR for
neutrinos:
\begin{enumerate}

\item{%
Using BR calculations we can probe
only part of the parameter space.
In the two generation case
three parameters describe the results of the
neutrino oscillation experiments
(see Eq.(\ref{simsec})): $\Delta m^2$,
$\theta_{ms}$ and $\theta_{md}$.
The BR calculation is sensitive only to one parameter,
the mixing angle that rotate
from the source basis
into the basis we are interested in.
For example,
the BR into final mass eigenstate is given by
${\rm BR}(\pi \to \mu \nu_1^m)=\sin^2\theta_{ms}$.}

\item{%
In order to compare BR calculations with experiments,
the dependence of the experimental results on
$\Delta m^2$ and $\theta_{md}$ has to be removed.
However, this cannot be done in a model independent way
since each kind of NP may contribute differently
to neutrino masses and to the detection process.
Therefore, experimental measurements of BR cannot be presented in a model
independent way.}

\item{Finally, there is a problem of definition.
Are the theoretical calculations and the
experimental bounds on rare decays as
${\rm BR}(\pi \to \mu \nu_e)$ \cite{Cooper,KARMEN},
${\rm BR}(K \to \mu \nu_e)$ \cite{Lyons}
and ${\rm BR}(\mu \to e \bar\nu_e \nu_\mu)$ \cite{Freedman,Krakauer,KARMEN}
related to the same quantities and therefore directly comparable?
Calculations
were done for neutrinos in the weak basis \cite{davidson},
and in the mass basis \cite{Moh,MPal}.
Experimental results are presented as bounds on the
relative appearance probability,
$\tilde{P}_{e\mu}^{exp}(i)=P_{e\mu}^{exp}/a_i$,
where $a_i$ is the relative weight of the relevant decay
mode in the neutrino beam.
In the $x \to 0$ limit they correspond to bounds on
the BR for neutrinos in the
detector basis, {\it e.g.} ${\rm BR}(\pi \to \mu \nu_e^d)$.
In the $x \to \infty$ limit, all the relevant angles are small, and
$\nu_1^m$ ($\nu_2^m$) couples mainly to the electron (muon). Then,
to first order in small angles,
electrons are detected when
$\nu_1^m$ is produced at the source, or when
$\nu_2^m$ creates an electron at the detector.
Therefore, the experimental results bound the sum
of the rare BR in the production process and the
ratio of the cross sections in
the detecting process, {\it e.g.}
${\rm BR}^{exp}(\pi \to \mu \nu_e)={\rm BR}(\pi \to \mu \nu_1^m)+
\sigma(\nu_2^m n \to ep)/\sigma(\nu_1^m n \to ep)$.
We see that the calculations cannot be directly compared with
the published experimental bounds.}

\end{enumerate}

We conclude: The notion of BR can be used to probe only
part of the parameter space. The BR
cannot be measured in a model independent way,
and comparisons of experimental and theoretical
results have to be done always very carefully,
paying attention to check that
the definitions are consistent.
Therefore, the notion of BR is not appropriate when searching for NP
in neutrino oscillation experiments, and the
formalism that we have developed here is preferable.

\section{examples}
We give now three examples of NP scenarios with
non-standard neutrino interactions
that can be probed
by current and near future neutrino oscillation experiments.
In each case we first
briefly present the model and the new neutrino interaction,
then we discuss the actual
experiment for probing the new interaction.
Then, we find the neutrino sector independent
experimental constraints on the strength on the new interaction,
and show how results of neutrino oscillation experiments
can put stronger bounds on it in
part of the neutrino sector parameter space.
For simplicity, we do not specify
the NP responsible for neutrino masses.

In the first example we consider the
Minimal Supersymmetric SM (MSSM) without R-Parity \cite{RSUSY,BGH}.
We consider a very simple case where the MSSM superpotential is
extended with only one extra term,
$\la_{123}[L^1_L,L^2_L]\bar E^3_R$,
where $L^i_L$ ($E^i_R$)
are lepton doublet (singlet) supermultiplets.
Via the exchange of the singlet charged scalar, $\tt=\tilde{E}^3_R$,
such a term gives rise to the
effective four fermion interaction
(in the weak basis) \cite{BGH}
\beq \label{cousn}
{\cal L}=G_N \left(\bar{\mu} \gamma_\lambda P_L e\right)
\left(\bar{\nu}^W_\mu \gamma^\lambda P_L \nu^W_e \right)\ +\ {\rm h.c.}\,,
\eeq
where $P_L=(1-\gamma_5)/2$
and
$G_N \sim {|\la_{123}|^2 / m_{\tt}^2}$.
(Recall: in the SM
${\cal L}_{SM}={\cal L}(\nu_\mu \leftrightarrow \nu_e,G_N \to G_F)$.)
Let us now consider the KARMEN experiment \cite{KARMEN}.
Muon anti-neutrinos are produced in muon decay,
$\mu^+ \to e^+ \nu_e^s \bar\nu_\mu^s$,
and electron
anti-neutrinos are searched through inverse
beta decay, $\bar \nu_e^d p \to e^+ n$.
Since $\tt$ couples only to leptons,
the detector basis coincides with the weak basis, but
the source basis is different.
Muon decay is mediated by $W$ or
$\tt$ exchange. In the weak basis, the $W$ diagram produces
$\bar{\nu}_\mu^W$, while the $\tt$ diagram produces $\bar{\nu}_e^W$.
The strongest neutrino sector independent bound on $G_N$ is obtained
from tests of universality in $\mu$ and $\beta$ decays \cite{FYo,BGH}.
{}From the lower bound \cite{PDG},
$\sum_{i=1}^3 |V_{ui}|^2> 0.995$, we get
\beq \label{boundmb}
\sin^2\theta_{sW} \sim
\left({G_N \over G_F}\right)^2 \lesssim 5 \times 10^{-3}.
\eeq
We like to show how the recent
90\% CL bound from KARMEN \cite{KARMEN}
\beq \label{karres}
P_{e \mu} \leq 3.1 \times 10^{-3},
\eeq
can be used to set stronger bounds on $G_N$ in part of the
neutrino sector parameter space.
We study two limiting cases. For massless neutrinos,
from Eq.(\ref{xzero}) we get the bound
\beq \label{rpxz}
\left({G_N \over G_F}\right)^2 \lesssim  3.1 \times 10^{-3},
\eeq
which is stronger than the bound (\ref{boundmb}).
In the large $\Delta m^2$ limit,
from Eq.(\ref{xinfty}) we get the combine bound
\beq \label{rpxin}
\left({G_N \over G_F}\right)^2 +
2\left({G_N \over G_F}\right)\theta_{mW} + 2\theta_{mW}^2  \lesssim
3.1 \times 10^{-3}.
\eeq

In the second example
we consider the minimal Left Right Symmetric (LRS) model \cite{LRS}.
In this model there is a Higgs triplet, $\Delta_L$,
with the Yukawa couplings to leptons (in the weak basis),
${\cal L}=f_{ij}L_i^T C i \tau_2 \Delta_L P_L L_j$,
where $L_i$ are the lepton doublets and $C$ is
the charge conjugation matrix. $\Delta_L^+$ exchange
leads to the effective
four fermion interaction in Eq.(\ref{cousn}) but with
$G_N \sim {|f_{11}f_{22}| / m_{\Delta_L^+}^2}$.
We again consider the KARMEN experiment.
Since $\Delta_L$ couples only to leptons,
the detector basis coincides with the weak basis, but
the source basis is different.
Muon decay is mediated by $W$ or
$\Delta_L^+$ exchange. In the weak basis, the $W$ diagram produces
$\bar{\nu}_\mu^W$, while the $\Delta_L^+$ diagram produces $\bar{\nu}_e^W$.
The strongest neutrino sector independent bound on $G_N$ is obtained
from tests of universality in $\mu$ and $\beta$ decays and
is given in Eq.(\ref{boundmb}).
Therefore,
the bounds (\ref{rpxz}) and (\ref{rpxin}) also hold for
the effective interaction arising from $\Delta_L^+$ exchange
in the minimal LRS model.

In the third example we
consider models with light leptoquarks (LQ) \cite{LQ}.
There are several types of LQ that can lead to sufficiently
large new neutrino interaction.
We concentrate on the $(I_3)_Y=(0)_{1/3}$
scalar LQ, $S$, which couples to fermions
(in the weak basis),
${\cal L}=\lambda_{ij} \bar{Q}^c_j i \tau_2 P_L L_i S$,
where $Q_i$ ($L_i$)
are the quark (lepton) doublets.
$S$ exchange leads to the effective four
fermion interaction (in the weak basis)
\beq \label{cousnn}
{\cal L}=G_N \left[(\bar{u} \gamma_\mu P_L d)
(\bar{\nu}^W_\tau \gamma_\mu P_L \mu)
+ (\bar{d} \gamma_\mu P_L u)
(\bar{\nu}^W_\mu \gamma_\mu P_L \tau)\right]\ +\  {\rm h.c.}\,
\eeq
with
$G_N \sim {|\la_{21}\la_{31}| / m_{LQ}^2}$. We assume $\la_{32} \ll \la_{31}$.
Therefore, LQ interactions involving strange quarks are negligible.
Let us consider the CHORUS, NOMAD and E803 experiments \cite{revexp}.
Muon neutrinos are produced in pion and kaon decays,
$\pi \to \mu \nu$ and $K \to \mu \nu$, and
tau neutrinos are searched through
$\nu n   \to \tau p$.
The new interaction (\ref{cousnn}) contributes the same
to pion decay and to the
detection process.
Therefore, the neutrino produced in pion decay is
$\nu_\mu^d$. This illustrates the above mentioned result:
had the neutrino beam been produced only
from pion decay, we could not probe
LQ exchange in those experiments.
However, the new interactions (\ref{cousnn})
do not contribute to kaon decay and
the neutrino produced in $K \to \mu \nu$ is a weak eigenstate
muon neutrino, $\nu_\mu^W$.
The strongest neutrino sector independent bound on $G_N$ is obtained
from the bound on
${\rm BR}(\tau \to \pi^0 \mu)$ \cite{davidson}
\beq
\left({G_N \over G_F}\right)^2 \approx
{{\rm BR}(\tau \to \pi^0 \mu)\over 2{\rm BR}(\tau \to \pi^- \nu)}.
\eeq
Using \cite{PDG}
${\rm BR}(\tau \to \pi^0 \mu) < 4.4 \times 10^{-5}$
and  ${\rm BR}(\tau \to \pi^- \nu) \approx 0.117$ we get
\beq \label{boundmbt}
\sin^2\theta_{dW} \sim
\left({G_N \over G_F}\right)^2 \lesssim 2 \times 10^{-4}.
\eeq
We like to show how the expected sensitivity of CHORUS and NOMAD,
$P_{\mu\tau}^{exp} \sim 10^{-4}$ and E803, $P_{\mu\tau}^{exp} \sim 10^{-5}$
\cite{revexp} can be used to probe LQ exchange in part of the
neutrino sector parameter space.
Since we concentrate on LQ that
can be tested only by neutrinos from kaon decay, we have
to use
the relative appearance probability of neutrinos from kaon decay,
where we expect
$\tilde{P}_{\mu\tau}^{exp}(K \to \mu \nu) = {\it few} \times
P_{\mu\tau}^{exp}$.
We learn that LQ exchange can be tested, probably at
CHORUS and NOMAD, and certainly by E803.
For example, assuming that the bound
$\tilde{P}_{\mu\tau}^{exp}(K \to \mu \nu) \lesssim 5 \times  10^{-5}$
will be achieved by E803.
For massless neutrinos,
Eq.(\ref{xzero}) gives
\beq \label{lqxz}
\left({G_N \over G_F}\right)^2 \lesssim  5 \times 10^{-5},
\eeq
which is stronger than the bound (\ref{boundmbt}).
In the large $\Delta m^2$ limit,
Eq.(\ref{xinfty}) gives
\beq \label{lqxin}
\left({G_N \over G_F}\right)^2 +
2\left({G_N \over G_F}\right)\theta_{mW} + 2\theta_{mW}^2  \lesssim
5 \times 10^{-5}.
\eeq

\section{Summary}
There are two important ways in which
physics beyond the SM can affect the neutrino sector:
It may give
non-vanishing neutrino masses, and it may modify
neutrino interactions. In this paper we showed how
neutrino oscillation experiments
probe both effects, and how they can be distinguished in some cases.
A distance-independent signal can arise from both effects. However,
an oscillation pattern can arise {\it only} when neutrinos are massive.
We study the condition on the relative strength of the non-standard
neutrino interactions
in order that it can be tested (see Eq.(\ref{npimpf})).
Thus, current experiments aimed to reach  a sensitivity of
the order $P_{ij}^{exp} \approx 10^{-4}$ \cite{revexp}
can typically probe new neutrino interactions arising from
physics at the 1 TeV scale.
There are several well motivated NP scenarios
that introduce non-diagonal couplings that can be probed in this way.
Higgs triplet exchange in Left-Right Symmetric models \cite{LRS},
Light leptoquarks exchange in various models \cite{LQ} and
super-particles exchange in Supersymmetric
models without R-parity \cite{RSUSY,BGH}.

Our results are of particular interest in light of the growing evidences
for physics beyond the SM in the neutrino sector,
in particular, the solar neutrino problem \cite{rev,Lan},
the atmospheric  neutrino deficit \cite{revatmo} and the
recent LSND result \cite{LSND}.
Those results cannot be
simultaneously accounted for
in a simple
three generation model, but only in models with more parameters.
Usually, it is suggested that those results are hints to
models with an extended neutrino sector \cite{exte}.
Alternatively, it might be that they
signal NP in neutrino interactions.
Such NP can be significant for
the MSW solution of the solar neutrino problem
\cite{FYo,mMSW}, for  atmospheric  neutrinos \cite{enrico}
and, as we have discussed, for
accelerator experiments.
A comprehensive analysis of all these experiments,
including possible NP, is needed.

\acknowledgements
I thank Enrico Nardi and Yossi Nir for many discussions and comments on the
manuscript, and Shmuel Nussinov and Nathan Weiss for helpful conversations.

{\tighten

}

\end{document}